\documentclass[10pt,a4paper]{article}
\usepackage[utf8]{inputenc}
\usepackage{amsmath}
\usepackage{amsfonts}
\usepackage{amssymb}
\usepackage[pdftex]{graphicx} 

\begin{document}

\section*{ }

\textbf{EXPONENTIAL OR POWER LAW? HOW TO SELECT A STABLE DISTRIBUTION OF PROBABILITY IN A A PHYSICAL SYSTEM} \\ \\

\noindent Andrea Di Vita, D.I.C.C.A., Università di Genova, Italy \\ 

\noindent $ andrea.divita@ansaldoenergia.com $ \\ \\

\noindent The 4th International Electronic Conference on Entropy and Its Applications  (ECEA 2017), 21 November--1st December 2017; Sciforum Electronic Conference Series, Vol. 4, 2017; doi:$ \mbox{10.3390/ecea-4-05009} $

\section{The problem}
\label{SEC1}
 
Usefulness of familiar, Gibbs' thermodynamics lies in its ability to provide predictions concerning systems at thermodynamical equilibrium with the help of no detailed knowledge of the dynamics of the system. The distribution of probabilities of the microstates in canonical systems described by Gibbs' thermodynamics is proportional to a Boltzmann exponential. 

No similar generality exists for those systems in steady, stable ('relaxed') state which interact with external world, which are kept far from thermodynamical equilibrium by suitable boundary conditions and where the probability distribution follows a power law. (Here we limit ourselves to systems where only Boltzmann-like or power-law-like distributions are allowed). Correspondingly, there is no way to ascertain whether the probability distribution in a relaxed state is Boltzmann-like or power-law-like, but via solution of the detailed equations which rule the dynamics of the particular system of interest. In other words, if we dub 'stable distribution function' distribution of probabilities of the microstates in a relaxed state, then no criterion exists for assessing the stability of a given probability distribution - Boltzmann-like or power-law-like - against perturbations. 

Admittedly, a theory exists - the so-called 'non-extensive statistical mechanics' \cite{TsallisORIG} \cite{TSALLISMENDESPLASTINO} \cite{Tsallis2} \cite{Abe} \cite{Tatsuaki} \cite{Umarov} - which extends the formal machinery of Gibbs' thermodynamics to systems where the probability distribution is power-law-like. Non-extensive statistical mechanics is unambiguously defined, once the value of a dimensionless parameter $ q $ is known; among other things, this value describes the slope of the probability distribution. If $ q \neq 1 $ then the quantity corresponding to the familiar Gibbs' entropy is not additive; Gibbs' thermodynamics and Boltzmann's distribution are retrieved in the limit $ q \rightarrow 1 $. Thus, if we know the value of $ q $ then we know if the distribution function of a stable, steady state of a system which interacts with external world is Boltzmann or power law, and, in the latter case, what its slope is like. Unfortunately, the problem is only shifted: in spite of the formal exactness of non-extensive statistical mechanics, there is no general criterion for estimating $ q $ - with the exception, again, of the solution of the equations of the dynamics. 

The aim of the present work is to find such criterion, for a wide class of physical sytems at least.
 
To this purpose, we recall that - in the framework of Gibbs' thermodynamics - the assumption of 'local thermodynamical equilibrium' (LTE) is made in many systems far from thermodynamical equilibrium, i.e. it is assumed that thermodynamical quantities like pressure, temperature etc. are defined withn a small mass element of the system and that these quantities are connected to each other by the same rules - like e.g. Gibbs-Duhem equation - which hold at true thermodynamical equilibrium. If, furthermore, LTE holds at all times during the evolution of the small mass element, then the latter satisfies the so-called 'general evolution criterion' (GEC), an inequality involving total time derivatives of thermodynamical quantities \cite{GlansdorffPrigogine}. Finally, if GEC holds for arbitrary small mass element of the system, then the evolution of the system as a whole is constrained; if such evolution leads a system to a final, relaxed state, then GEC puts a constraint on relaxation. 

Straightforward generalization of these results to the non-extensive case $ q \neq 1 $ is impossible. In this case, indeed, the very idea of LTE is scarcely useful: the $ q \neq 1 $ entropy being a non-additive quantity, the entropy of the system is not the sum of the entropies of the small mass elements the system is made of, and no constraint on the relaxation of the system as a whole may be extracted from the thermodynamics of its small mass elements of the system. (For mathematical simplicity, we assume $ q $ to be uniform  across the system). 

All the same, an additive quantity exists which is monotonically increasing with the entropy (and achieves therefore a maximum if and only if the $ q \neq 1 $ entropy is maximum) and which reduces to Gibbs' entropy as $ q \rightarrow 1 $. Thus, the $ q \neq 1 $ case may be unambiguously mapped onto the corresponding Gibbs' problem \cite{VivesandPlanes}, and all the results above still apply. 
As a consequence, a common criterion of stability exists for relaxed states for both $ q = 1 $ and $ q \neq 1 $. The class of perturbations which the relaxed states satisfying such criterion may be stable against include perturbations of $ q $. 

We review some relevant results of non-extensive thermodynamics in Sec. \ref{SEC2} . The role of GEC and its consequences in Gibbs' thermodynamics is discussed in Sec. \ref{SEC3} . Sec. \ref{SEC4} discusses generalization of the results of Sec. \ref{SEC3} to the $ q \neq 1 $ case. Sec. \ref{SEC5} shows application to a simple toy model. We apply the results of Sec. \ref{SEC5} to a class of physical problems in Sec. \ref{SEC5bis} . Conclusions are drawn in Sec. \ref{SEC6}. Entropies are normalized to Boltzmann's constant $ k_{B} $.
  
\section{Power-law vs. exponential distributions of probability}
\label{SEC2}

For any probability distribution $ \left\lbrace p_{k} \right\rbrace $ defined on a set of $ k = 1, \ldots W $ microstates of a physical system, the following quantity  \cite{TsallisORIG}

\begin{equation}
\label{def0}
S_{q} = - \sum_{k} \left( p_{k} \right)^q \ln_{q} p_{k}
\end{equation}

is defined, where $ \ln_{q} x \equiv \frac{x^{1-q}-1}{1-q}$ is the inverse function of $ \exp_{q} \left( x \right) \equiv 
\left[ 1 + \left( 1 - q \right) x \right]^{\frac{1}{1-q}} $ and $ q > 0 $. 

For an isolated (microcanonical) system, constrained maximization of $ S_{q} $ leads to $ p_{k} = \frac{1}{W} $ for all $k$'s and to $ S_{q} = \ln_{q} \left( W \right) $, the constraint being given by the normalization condition 
$ \sum_{k} p_{k} = 1 $.

For non-isolated systems \cite{TSALLISMENDESPLASTINO} \cite{VivesandPlanes}, some ($ i = 1, \ldots M $) quantities - e.g. energy, number of particles etc. - whose values $ x_{ik} $ label the $ k $-th microstate and which are additive constants of motion in an isolated system become fixed only on average (the additivity of a quantity signifies that, when the amount of matter is changed by a given factor, the quantity is changed by the same factor \cite{Landau}). Maximization of $ S_{q} $ with the normalization condition $ \sum_{k} p_{k} = 1 $ and the further $ M $ constraints 
$ X_{i} \equiv x_{ik} P_{\left( q \right) k} = $ const. (each with Lagrange multiplier $ Y_{i} $ and $ P_{\left( q \right) k} \equiv \frac{\left( p_{k} \right) ^q}{\sum_{k} \left( p_{k} \right) ^q} $; repeated indices are summed here and below)  
leads to $ S_{q} = \ln_{q} \left( Z_{q} \right)$, $ Z_{q} = \sum_{k} \exp_{q} \left( - Y_{i} F_{ik} \right) $, $ F_{ik} \equiv \frac{x_{i} \left( k \right) - X_{i}}{\sum_{k} \left( p_{k} \right) ^q}$ and to the following, power-law-like probability distribution:

\begin{equation}
\label{def1}
p_{k} = \dfrac{\exp_{q} \left( -  Y_{i} F_{ik} \right) }{Z_{q}}
\end{equation}

Remarkably, equations (53) of \cite{TSALLISMENDESPLASTINO} and (6) of \cite{Tsallis2} show that suitable rescaling of the $ Y_{i} $'s allows us to get rid of the denominator $ \sum_{k} \left( p_{k} \right) ^q $ in the $ F_{ik} $'s and to make all computations explicit - in the case $ M = 1 $ at least. Finally, if we apply a quasi-static transformation to a $ S_{q} = \max $ state then:

\begin{equation}
\label{def2}
d S_{q} = Y_{i} d X_{i}
\end{equation}

If $ q \rightarrow 1 $ then \eqref{def0} and \eqref{def1} lead to Gibbs' entropy $ S_{q = 1} = - \sum_{k}p_{k}\ln p_{k} $ and to Boltzmann's, exponential probability distribution respectively. 

Many results of Gibbs' thermodynamics still hold if $ q \neq 1 $. For example, a Helmholtz' free energy $ F_{q} $ still links $ S_{q} $ and $ Z_{q} $ the usual way  \cite{TSALLISMENDESPLASTINO} \cite{Abe}. Moreover, if two physical systems $ A' $ and $ A'' $ are independent (in the sense that the probabilities of $ A' $ + $ A'' $ factorize into those of $ A' $ and of $ A'' $) then we may still write for the averaged values of the additive quantities \cite{TSALLISMENDESPLASTINO}

\begin{equation}
\label{add}
X_{i} \left( A' + A'' \right) = X_{i} \left( A' \right) + X_{i} \left( A'' \right)
\end{equation}

Generally speaking, however, \eqref{add} does not apply to $ S_{q} $, which satisfies:

\begin{equation}
\label{nonadd}
S_{q} \left( A' + A'' \right) = S_{q} \left( A' \right) + S_{q} \left( A'' \right) + \left( 1 - q \right) S_{q} \left( A' \right) S_{q} \left( A'' \right)
\end{equation}

\section{$ q = 1 $}
\label{SEC3}

Equations \ref{add} and \eqref{nonadd} are relevant when it comes to discuss stability of the system $ A' $ + $ A'' $ against perturbations localized inside an arbitrary, small subsystem $ A' $. (It makes still sense to investigate the interaction of $ A' $ and $ A'' $ while dubbing them as 'independent', as far as the internal energies of $ A' $ and $ A'' $ are large compared with their interaction energy \cite{Landau}). Firstly, we recollect some results concerning the well-known case $ q = 1 $; then, we investigate the $ q \neq 1 $ problem. 

To start with, we assume that $ M = 2 $; generalization to $ M \neq 2 $ follows. We are free to choose $ x_{1k} $ and $ x_{2k} $ to be the energy and the volume of the system in the $ k $-th microstate respectively. Then $ Y_{1} = \frac{\partial S_{q}}{\partial X_{1}} = \beta \sum_{k} \left( p_{k} \right) ^q $ and $ Y_{2} = \frac{\partial S_{q}}{\partial X_{2}} = \beta p \sum_{k} \left( p_{k} \right) ^q $ \cite{Abe} with $ \beta \equiv \frac{1}{k_{B} T} $ and where $ T = k_{B}^{-1} \left( \frac{\partial S_{q = 1}}{\partial U} \right)_{V}^{-1}$, $ p = - \left( \frac{\partial F_{q = 1}}{\partial V} \right)_{T}$, $ U \equiv \lim_{q \rightarrow 1} X_{1} $ and $ V \equiv \lim_{q \rightarrow 1} X_{2}$ are the familiar absolute temperature, pressure, internal energy and volume respectively. In the limit $ q \rightarrow 1 $ we have $ \sum_{k} \left( p_{k} \right) ^q  = 1 + (1 - q ) S_{q} \rightarrow 1 $, $ X_{1} $, the familiar thermodynamical relationships $ \left( \frac{\partial S_{q = 1}}{\partial U} \right)_{V} = \beta $ and $ \left( \frac{\partial S_{q = 1}}{\partial V} \right)_{T} = \beta p $ are retrieved, and \eqref{def2} is just a simple form of the first principle of thermodynamics.

Since $ q = 1 $, \eqref{nonadd} ensures additivity of Gibbs' entropy. We assume $ A' $ to be is at thermodynamical equilibrium with itself, i.e. to maximize $ S_{q = 1} \left( A' \right) $ (LTE). We allow $ A' $ to be also at equilibrium with the rest $ A'' $ of the system $ A' $ + $ A'' $, until some small, external perturbation occurs and destroys such equilibrium. The first principle of thermodynamics and the additivity of $ S_{q = 1} $ lead to Le Chatelier's principle \cite{Landau}. In turn, such principle leads to 2 inequalities, $ \left( \frac{\partial S}{\partial T} \right)_{V} > 0 $ and $ \left( \frac{\partial p}{\partial V} \right)_{T} < 0$. States in which such inequalities are not satisfied are unstable. 

Let us introduce the volume $ dV $, the mass density $ \rho $ and the mass $ \rho dV $ of $ A' $. (Just like $ \rho $, here and in the following we refer to the value of the generic physical quantity $ a $ at the center of mass of $ A' $ as to 'the value of $ a $ in $ A' $ '; this makes sense, provided that $ A' $ is small enough). Together with the additivity of Gibbs' entropy, arbitrariness in the choice of $ A' $ ensures that $ S_{q} \left( A' + A'' \right) = \int dV \rho s $ where $ S_{q} \left( A' + A'' \right) $ and $ s $ are Gibbs' entropy of the whole system $ A' + A'' $ and Gibbs' entropy per unit mass respectively; here and in the following, integrals are extended to the whole system $ A' $ + $ A'' $. The internal energy per unit mass $ u $ and the volume per unit mass ($ = \frac{1}{\rho} $) may similarly be introduced, as well as the all the quantities per unit mass corresponding to all the $ X_{i} $'s which satisfy \eqref{add}. Inequalities $ \left( \frac{\partial S}{\partial T} \right)_{V} > 0 $ and $ \left( \frac{\partial p}{\partial V} \right)_{T} < 0$ lead to $ \left( \frac{\partial s}{\partial T} \right)_{V} > 0 $ and $ \left( \frac{\partial p}{\partial \rho} \right)_{T} > 0$ respectively.

We relax the assumption $ M = 2 $. If $ A' $ contains particles of $ h = 1, \ldots N $  chemical species, each with $ N_{h} $ particles with mass $ m_{h} $ and chemical potential $ \mu_{h} $, then $ N $ degrees of freedom add to the 2 degrees of freedom $ U $ and $ V $, i.e. $ M = N + 2 $. In the $ k$-th microstate, $ x_{h + 2,k} $ is the number of particles of the $ h$-th species. In analogy with $ U $ and $ V $, we write $ N_{h} = \lim_{q \rightarrow 1} X_{h + 2}$. Starting from this $ M $ additive quantities, different $ M $-ples of coordinates (thermodynamical potentials) may be selected with the help of Legendre transforms. LTE implies minimization of Gibbs' free energy $ F_{q = 1} + pV = \mu_{h} N_{h}$ at constant $ T $ and $ p $. As for quantities per unit mass, this minimization leads to the inequality $ \left( \frac{\partial \mu_{h}^{o}}{\partial c_{j}} \right)_{p,T} dc_{h} dc_{j} \geq 0$ \cite{PrigogineDefay} where $ \mu_{h}^{o} = \frac{\mu_{h}}{m_{h}} $, $ c_{j} \equiv \frac{N_{j} m_{j} }{\sum_{h} N_{h} m_{h}} $, $ j = 1, \ldots N $. Identity $ \sum_{h} c_{h} = 1$ reduces $ M $ by 1. With this proviso, we conclude that validity of LTE in A requires:

\begin{equation}
\label{stab}
\left( \frac{\partial s}{\partial T} \right)_{V,N} > 0 
\quad ; \quad
\left( \frac{\partial p}{\partial \rho} \right)_{T,N} > 0
\quad ; \quad
\left( \frac{\partial \mu_{h}^{o}}{\partial c_{j}} \right)_{p,T} dc_{h} dc_{j} \geq 0
\end{equation}

where $ \left( \right)_{N} $ means that all $ c_{h}$'s are kept fixed, and $ \geq $ is replaced by $ = $ only for $ dc_{h} = 0$. The $ 1^{st} $, $ 2^{nd} $ and $ 3^{rd} $ inequality in \eqref{stab} refer to thermal, mechanical and chemical equilibrium respectively.

Remarkably, \eqref{stab} contains information on $ A' $ only; $ A'' $ has disappeared altogether. Thus, if we allow $ A' $ to change in time (because of some unknown, physical process occurring in $ A'' $, which we are not interested in at the moment) but we assume that LTE remains valid at all times within $ A' $ followed along its center-of-mass motion ($ \textbf{v} $ being the velocity of the center-of-mass), then \eqref{stab} remains valid in $ A' $ at all times. In this case, all relationships among total differentials of thermodynamic quantities - like e.g. Gibbs-Duhem equation - remain locally valid, provided that the total differential $ da $ of the generic quantity $ a $ is $ da = \frac{da}{dt}d $ where $ \frac{da}{dt} = \frac{\partial a}{\partial t} + \textbf{v} \cdot \nabla a $. Thus, \eqref{stab} leads to the so  called 'general evolution criterion' (GEC) \cite{GlansdorffPrigogine} \cite{Di Vita}

\begin{equation}
\label{GEC}
\dfrac{dT^{-1}}{dt} \dfrac{d \left( \rho u \right)}{dt} -
\rho \sum_{h} \dfrac{d\left(\mu_{h}^{o}T^{-1}\right)}{dt} \dfrac{d c_{h}}{dt} - 
\left[
\rho^{-1}T^{-1}\dfrac{dp}{dt}
+ \left( u + \rho^{-1}p \right) \dfrac{dT^{-1}}{dt}
\right]
\dfrac{d\rho}{dt}
\leq 0
\end{equation}

No matter how erratic the evolution of $ A' $ is, if LTE holds within $ A' $ at all times then the (by now) time-dependent quantities $ T \left( t \right) $, $ \rho \left( t \right) $ etc. satisfy \eqref{GEC} at all times. 

GEC is relevant to stability. By 'stability' we refer to the fact that, according to Einstein's formula \cite{Landau}, deviations from the $ S_{q = 1} = \max$ state which lead to a reduction of Gibbs' entropy ($ \Delta S_{q = 1} < 0$) have vanishing small probability $ \propto \exp \left( \Delta S_{q = 1} \right) $. Such deviations can e.g. be understood as a consequence of an internal constraint which causes the deviation of the system from the equilibrium state, or as a consequence of contact with an external bath which allows changes in parameters which would be constant under total isolation. Let us characterize  this deviation by a parameter $ \kappa $ which vanishes at equilibrium. Einstein's formula implies that small $ \kappa $ fluctuations near the configuration which maximizes $ S_{q = 1} $ are Gaussian distributed with variance $ \left( \frac{\partial^2 S_{q = 1} }{\partial \kappa^2}\right)^{-1} $.

Correspondingly, as far as $ A' $ is at LTE deviations of the probability distribution $ \left\lbrace p_{k} \right\rbrace $ from Boltzmann's exponential distribution are also extremely unlikely. As $ A' $ evolves, the instantaneous values of the $ X_{i} $'s and the $ Y_{i} $'s may change, but if LTE is to hold then the shape of $ p_{k} $ remains unaffected. For example, $ T $ may change in time, but the probability of a microstate with energy $ E $ remains $ \propto \exp \left( - \beta E \right) $. Should Boltzmann's distribution becomes unstable at any time - i.e., should any deviation of $ p_{k} $ from Boltzmann's distribution ever fail to fade out - then LTE too should be violated, and \eqref{GEC} cease to hold. Then, we conclude that if $ p_{k} $ remains Boltzmann-like in $ A' $ at all times then \eqref{GEC} remains valid in $ A' $ at all times.

As for the evolution of the whole system $ A' $ + $ A'' $ as a whole, if LTE holds everywhere throughout the whole system at all times then \eqref{GEC} too holds everywhere at all times. In particular, let the whole system $ A' $ + $ A'' $ evolve towards a final, relaxed state, where we maintain - as a working hypothesis - that the word 'steady' makes sense, possibly after time-averaging on some typical time scales of macroscopic physics. Since LTE holds everywhere at all times during relaxation, \eqref{GEC} puts a constraint on relaxation everywhere at all times; as a consequence, it provides us with information about the relaxed state as well. In the following, we are going to show that some of the above result find its counterpart in the $ q \neq 1 $ case.

\section{$ q \neq 1 $}
\label{SEC4} 

If $ q \neq 1 $ then \eqref{nonadd} ensures that $ S_{q} $ is not additive; moreover, it is not possible to find a meaningful expression for $ s $ such that $ S_{q \neq 1} \left( A' + A'' \right) = \int dV \rho s $, and the results of Sec. \ref{SEC3} fail to apply to $ S_{q} $ (see Appendix \ref{AA}). All the same, even if $ q \neq 1 $ the quantity 

\begin{equation}
\label{Sc}
\widehat{S_{q}} \equiv \dfrac{\ln \left( 1 + \left( 1 - q \right) S_{q} \right)}{1 - q}
\end{equation}

is additive and satisfies the conditions $ \lim_{q \rightarrow 1} \widehat{S_{q}} = S_{q = 1} $ and 

\begin{equation}
\label{monotonicdependence}
\dfrac{d \widehat{S_{q}} }{dS_{q}} > 0
\end{equation}

so that $ \widehat{S_{q}} = \max $ if and only if $ S_{q} = \max $ \cite{TsallisORIG} \cite{VivesandPlanes} \cite{Abe} \cite{Marino}. Then, a power-law-like distribution \eqref{def1} corresponds to $ \widehat{S_{q}} = \max $. Moreover, the additivity of $ \widehat{S_{q}} $ makes it reasonable to wonder whether a straightforward, step-by-step repetition of the arguments of Sec. \ref{SEC3} leads to their generalization to the $ q \neq 1 $ case. 
When looking for an answer, we are going to discuss each step separately.

First of all, the choice of the $ x_{ik} $'s does not depend on the actual value of $ q $; then, the $ X_{i} $'s  are unchanged, and \eqref{add} still holds as it depends only on the averaging procedure on the $ p_{k} $'s. As anticipated, \eqref{def1} corresponds to a maximum of $ \widehat{S_{q}} $, and we replace \eqref{nonadd} with

\begin{equation}
\label{add2}
\widehat{S_{q}} \left( A' + A'' \right) = \widehat{S_{q}} \left( A' \right) + \widehat{S_{q}} \left( A'' \right)
\end{equation}

Since we are interested in probability distributions which maximize $ \widehat{S_{q}} $, hence $ S_{q} $, we are allowed to invoke equations (11) and (12) of \cite{VivesandPlanes} and to write the following generalization of \eqref{def2}: 

\begin{equation}
\label{def3}
d \widehat{S_{q}} = \widehat{Y_{i}} d X_{i}
\end{equation}

\begin{equation}
\label{def4}
\widehat{Y_{i}} = \dfrac{Y_{i}}{1 + \left( 1 - q \right) S_{q}}
\end{equation}

Once again, we start with $ M = 2 $ and choose $ x_{1k} $ and $ x_{2k} $ to be the energy and the volume of the system in the $ k $-th microstate respectively. Together, \eqref{def3}, \eqref{def4} and the identity $ \sum_{k} \left( p_{k} \right) ^q  = 1 + (1 - q ) S_{q} $ give 
$ \widehat{Y_{1}} = 
\frac{\partial \widehat{S_{q}}}{\partial X_{1}} = 
\frac{1}{1 + \left( 1 - q \right) S_{q}}\frac{\partial S_{q}}{\partial X_{1}} = 
\beta \frac{\sum_{k} \left( p_{k} \right) ^q}{\sum_{k} \left( p_{k} \right) ^q} = 
\beta $ 
and 
$ \widehat{Y_{2}} = 
\frac{\partial \widehat{S_{q}}}{\partial X_{2}} = 
\frac{1}{1 + \left( 1 - q \right) S_{q}}\frac{\partial S_{q}}{\partial X_{2}} = 
\beta p \frac{\sum_{k} \left( p_{k} \right) ^q}{\sum_{k} \left( p_{k} \right) ^q} = 
\beta p $, i.e. we retrieve the usual temperature and pressure of the $ q = 1 $ case \cite{Marino}. 

At last, \eqref{add} and \eqref{add2} allow us to repeat step-by-step the proof of \eqref{stab} and of \eqref{GEC}, provided that LTE now means that $ A' $ is in a state which corresponds to a maximum of $ \widehat{S_{q}} $. This way, we draw the conclusion that GEC takes exactly the same form \eqref{GEC} even if $ q \neq 1 $. In detail, we have shown that both $ T $, $ p $, $ X_{1} $ and $ X_{2} $ (i.e. $ U $ and $ V $) are unchanged; the same holds for $ u $ and $ \frac{1}{\rho} $. The 2$ ^{nd} $ inequality in \eqref{stab} remains unchanged: indeed, this is equivalent to say that the speed of sound remains well-defined in a $ q \neq 1 $ system - see e.g. \cite{Khuntia}. Admittedly, both the entropy per unit mass and the chemical potentials change when we replace $ S_{q = 1} $ with $ \widehat{S_{q}} $. However, $ \left( \frac{ \partial \widehat{S_{q}}}{\partial  T} \right)_{V , N} = 
\frac{d \widehat{S_{q}} }{dS_{q}} \left( \frac{ \partial S_{q} }{\partial  T} \right)_{V , N} $ 
has the same sign of $ \left( \frac{ \partial      S_{q} }{\partial  T} \right)_{V , N} $
because $ \frac{d \widehat{S_{q}} }{dS_{q}} > 0 $ and
$ \left( \frac{ \partial S_{q} }{\partial  T} \right)_{V , N} $
( $ \propto $ a specific heat) is $ > 0 $ \cite{Tatsuaki}. Thus, the 1$ ^{st} $ inequality in \eqref{stab} still holds because of the additivity of $ \widehat{S_{q}} $. Finally, maximization of Gibbs' free energy at fixed $ T $ and $ p $ follows from maximization of $  \widehat{S_{q}} $ as well as from \eqref{add} and \eqref{add2}, and the 3$ ^{rd} $ inequality in \eqref{stab} remains valid even if the actual values of the $ \mu_{h} $'s may be changed. 

Even the notion of stability remains unaffected. Equation (18) of \cite{VivesandPlanes}  generalizes Einstein's formula to $ q \neq 1 $ and ensures that strong deviations from the maximum of $ \widehat{S_{q}} $ are exponentially unlikely. As a further consequence of generalized Einstein's formula, if the deviation is characterized by a parameter $ \kappa $ which vanishes at equilibrium, then equation (21) of \cite{VivesandPlanes} ensures that small $ \kappa $ fluctuations near the configuration which maximizes $ \widehat{S_{q}} $ are Gaussian distributed with variance $ \left( \frac{\partial^2 \widehat{S_{q}} }{\partial \kappa^2}\right)^{-1} $ (and $ q \neq 1 $ fluctuations may be larger than $ q = 1 $ fluctuations). 

In spite of \eqref{nonadd}, \eqref{monotonicdependence} allows us to extend some of our results to $ S_{q} $. We have seen that configurations maximizing $ \widehat{S_{q}} $ maximize also $ S_{q} $ (where $ S_{q} $ is considered for the whole system $ A' $ + $ A'' $). Analogously, \eqref{monotonicdependence} implies $ \left( \frac{\partial^2 \widehat{S_{q}} }{\partial \kappa^2}\right)^{-1} \propto \left( \frac{\partial^2 S_{q} }{\partial \kappa^2}\right)^{-1} $. Given the link between $ S_{q} $ and \eqref{def1}, we may apply step-by-step our discussion of Botzmann's distribution to power-law distributions. By now, the role of $ \widehat{S_{q}} $ is clear: it acts as a dummy variable, whose additivity allows us to extend our discussion of Boltzmann's distribution to power-law distributions in spite of the fact that $ S_{q} $ is not additive. 

Our discussion suggests that if relaxed states exist, then thermodynamics provides a common description of relaxation regardless of the actual value of $ q $. As a consequence, thermodynamics may provide information about the relaxed states which are the final outcome of relaxation. Since relaxed states are stable against fluctuations and are endowed with probability distributions of the microstates, such information involves stability of these probability distributions against fluctuations. Since thermodynamics provides information regardless of $ q $, such information involves Boltzmann exponential and power-law distributions on an equal footing. We are going to discuss such information in depth for a toy model in Sec. \ref{SEC5}. In spite of its simplicity, the structure of its relaxed states are far from trivial. 

Below, it turns to be useful to define the following quantities. In the $ q = 1 $ case we introduce the contribution $ \Pi_{q = 1} $ to $ \frac{dS_{q = 1; A' + A''}}{dt} $ of the irreversible processes occurring in the bulk of the whole system $ A' + A'' $ ($ \Pi_{q = 1} $ is often referred to as $ \frac{d_{i}S}{dt} $ in the literature); by definition, such processes raise $ S_{q = 1} $ by an amount $ dt \cdot \Pi_{q = 1} $ in a time interval $ dt $. During relaxation, $ \Pi_{q = 1} $ is a function of time $ t $, and $ \Pi_{q = 1} \left( t \right) $ is constrained by \eqref{GEC}. A straightforward generalization of $ \Pi_{q = 1} $ to $ q \neq 1 $ is $ \widehat{\Pi_{q}} $, where $ dt \cdot \widehat{\Pi_{q}} $ is the growth of $ \widehat{S_{q}} $ due to irreversible processes in the bulk; $ \widehat{\Pi_{q}} \left( t \right) $ is constrained by the $ q \neq 1 $ version of \eqref{GEC} in exactly the same way of the $ q = 1 $ case. Finally, it is still possible to define $ \Pi_{q} $ such that the irreversible processes occurring in the bulk of the whole system $ A' + A'' $ raise $ S_{q} $ by an amount $ dt \cdot \Pi_{q} $. As usual by now, $ \lim_{q \rightarrow 1} \widehat{\Pi_{q}} = \lim_{q \rightarrow 1} \Pi_{q} = \Pi_{q = 1} $ and $ \Pi_{q} = \frac{d \widehat{S_{q}}}{dS_{q}} \cdot \widehat{\Pi_{q}} $. We provide an explicit epxression for $ \Pi_{q} $ in our toy model below.

\section{A toy model}
\label{SEC5}  

\subsection{A simple case.} The discussion of Sec. \ref{SEC4} does not rely on a particular choice of the $ X_{i} $'s, as the latter may be changed via Legendre transforms and their choice obviously leaves the actual value of amount of heat produced in the bulk unaffected. Moreover, it holds regardless of the actual value of $ M $ as $ dc_{h} = 0 $. As an example, we may think e.g. of a $M = 1$ system with just one chemical species, the volume $x_{2k}$ of the system in the $k$-th microstate is fixed and the energy $x_{1k}$ of the $k$-th microstate may change because of exchange of heat with the external world. Finally, our discussion is not limited to three-dimensional systems. We focus on a toy model with just 1 degree of freedom, which we suppose to be a continuous variable (say, $ x $) for simplicity so that we may replace $ p_{k} $ with a distribution function $ P \left( x , t \right) $ satisfying the normalization condition $ \int d\mbox{x} P = 1 $ at all times. We do not require that $ x $ retains its original meaning of energy: it may as well be the position of a particle. Here and below, integrals are performed on the whole system unless otherwise specified. The $ x $ runs within a fixed 1-D domain, which acts as a constant $ V $. In our toy model, the impact of a driving force $ A \left( x \right) $ is contrasted by a diffusion process with constant and uniform diffusion coefficient $ D = \alpha_{D} T > 0$. Following \cite{Casas}, we write the equation in $ P \left( x , t \right) $ in the form of a non linear Fokker Planck equation:

\begin{equation}
\label{FP}
\dfrac{\partial P}{\partial t} + \dfrac{\partial J}{\partial x} = 0
\quad ; \quad 
J = \dfrac{1}{\eta} \left( A P - D q' P^{q' - 1} \dfrac{\partial P}{\partial x} \right)
\end{equation}

Here $ \eta $ is a constant, effective friction coefficient, $ q' \equiv 2 - q $, and we drop the dependence on both $ x $ and $ t $ for simplicity here and in the following, unless otherwise specified. 

%
The value of $ q $ is assumed to be both known and constant in \eqref{FP}. Furthermore, if the value of $ q $ is known and a relaxed state exists, then \eqref{FP} describes relaxation. Now, if we allow $ q $ to change in time ($ q = q \left( t \right) $) much more slowly than the relaxation described by \eqref{FP}, then the evolution of the system is a succession of relaxed states. 
Sec. \ref{SEC4}. 
Unfortunately, available, GEC-based stability criteria \cite{Di Vita} are useless, as they have been derived for perturbations at constant $ q $ only. 
In order to solve this conundrum, and given the fact that the evolution of the system is a succession of relaxed states, we start with some information about such states. 

\subsection{Relaxed states.} In relaxed states $ \frac{\partial}{\partial t}\equiv 0 $ and \eqref{FP} implies 

\begin{equation}
\label{uniformJ}
\dfrac{\partial J}{\partial x} = 0
\end{equation}

where the value of $ J $ depends on the boundary conditions (e.g. the flow of $ P $ across the boundaries). In particular, the probability distribution \cite{Wedemann}:

\begin{equation}
\label{Pq}
P_{J = 0, q} \propto \exp_{q} \left( \int ^{x} d\mbox{x'} \dfrac{A}{D} \right)
\end{equation}

solves \eqref{FP} if and only if $ J = 0$ everywhere, i.e. if and only if the quantity   

\begin{equation}
\label{Piq}
\Pi_{q'} = \dfrac{\eta}{D} \int d\mbox{x} \dfrac{\vert J \vert^2}{P} 
\end{equation}

vanishes. The solution \eqref{Pq} is retrieved in applications - see e.g. equation (6) in \cite{Haubold} and equation (2) of \cite{Ribeiro}. The proportionality constant in the R.H.S. of equation \eqref{Pq} is fixed by the normalization condition $ \int d\mbox{x} P = 1$. The quantity $ \Pi_{q'} $ in \eqref{Piq} is the amount of entropy \cite{Casas}

\begin{equation}
\label{Sq}
S_{q'} = \int d\mbox{x} \dfrac{P - P^{q'}}{q' - 1}
\end{equation}

produced per unit time in the bulk; even if $ \frac{\partial}{\partial t} \neq 0 $ equations (10), (11), (42) and (44) of \cite{Casas} give:

\begin{equation}
\label{balanceofSq}
\Pi_{q'} - \dfrac{dS_{q'}}{dt} = \dfrac{1}{D} \int d\mbox{x} A J
\end{equation}

and in the R.H.S. of \eqref{balanceofSq} \textit{one identifies the entropy flux, representing the exchanges of entropy between the system and its neighborhood} per unit time, in the words of \cite{Casas}. In relaxed states such amount is precisely equal to $  \Pi_{q'} $ because $ \frac{dS_{q'}}{dt} = 0 $.

Admittedly, \eqref{Piq} and \eqref{Sq} deal with $ q' $, rather than with $ q $. In contrast, it is $ q $ which appears in \eqref{Pq}. Replacing $ q $ with $ q' $ is equivalent to replace $ 1 - q $ with $ q - 1 $; however, the duality $ q \leftrightarrow q' $ of non-extensive statistical mechanics - see Sec. 2 of \cite{Wada} and Sec. 6 of \cite{Naudts} - ensures that no physics is lost this way ($ S_{q'} $ replaces $ S_{q} $, etc.). Following Sec. III of \cite{Borland}, we limit ourselves to $ q < 2 $ for $ D > 0 $; the symmetry $ q \longleftrightarrow q' $ allow us therefore to focus further our attention on the interval $ 0 < q \leq 1 $ \cite{VivesandPlanes}. Not surprisingly, if $ q \rightarrow 1 $ then $ q' \rightarrow 1 $ and \eqref{Pq} and \eqref{Sq} reduce to Boltzmann's exponential (where $ \int ^{x} d\mbox{x'} \frac{A}{D} $ corresponds to $- \beta U $) and to Gibbs' entropy respectively.

At a first glance, $ S_{q'} $ is defined in two different ways, namely \eqref{def0} and \eqref{Sq}. However, identities $ \sum_{k} p_{k} = 1 $ and $ \int d\mbox{x} P = 1 $ allow \eqref{def0} and \eqref{Sq} to agree with each other, provided that we identify $ \sum_{k} \left( p_{k} \right) ^{q'} $ and $ \int  d\mbox{x} P^{q'} $; according to \eqref{Pq}, this is equivalent to a rescaling of $ \alpha_{D} $ and $ x $. Comparison of \eqref{def0} and \eqref{Sq} explains why it is not possible to find a meaningful expression for the entropy per unit mass $ s $ unless $ q \rightarrow 1 $ - see Appendix \ref{AA}.

According to \cite{Casas}, a H-theorem exists for \eqref{FP} even if $ J \neq 0 $, as far as $ \lim_{x \vert \rightarrow \infty} P = 0$, $ \lim_{x \vert \rightarrow \infty} \frac{dP}{dx} = 0$ and $ A $ is 'well-behaved at infinity', i.e. $ \lim_{x \vert \rightarrow \infty} AP = 0$; relaxed states minimize a suitably defined Helmholtz' free energy. 'Equilibrium' ($ S_{q} = \max $) occurs \cite{TSALLISMENDESPLASTINO} \cite{Borland} \cite{Wedemann} when \eqref{Pq} holds, i.e. for $ \frac{\partial}{\partial t} \equiv 0 $, $ J \equiv 0 $ and $ \Pi_{q'} = 0$; boundary conditions may keep the relaxed system 'far from equilibrium' ($ \frac{\partial}{\partial t} \equiv 0 $, $ J \neq 0 $, $ \Pi_{q'} > 0 $).

If $ J = 0 $ then \eqref{Piq} and \eqref{balanceofSq} ensure that no exchange of entropy between the system and its neighborhood occurs and that $ \Pi_{q'} = 0 $ regardless of $q'$. The probability distribution in the relaxed state of isolated systems is a Boltzmann's exponential. It is the the interaction with the external world (i.e., those boundary conditions which keep $J$ far from $ 0 $) which allows the probability distribution in the relaxed state of non-isolated systems to differ from a Boltzmann's exponential.

If $ J \neq 0 $ then the solution of \eqref{FP} in relaxed state satisfies the H-theorem quoted above. In the following we are going to discuss the 'weak dissipation' limit of small (but not vanishing) $ \vert J \vert $, which corresponds to weakly dissipating systems as \eqref{Piq} gives $ \Pi_{q'} = O \left( \vert J \vert^2 \right) $. 

\subsection{Weak dissipation.} The definition of $ J $ and \eqref{Piq} show how $ \Pi_{q'} $ depends on $ q' $. We may write this dependence more explicitly in the weak dissipation limit. We show in Appendix \ref{BB} that: 

\begin{equation}
\label{Taylor}
\Pi_{z} = \Pi_{z = 0} + \sum _{n = 1} ^{\infty} a_{n} z^n
\end{equation}

\begin{equation}
\label{coeffs}
a_{n} = \dfrac{\left( -1 \right)^{n - 1}\left( 2 J \right)}{\left( n - 1 \right)!} 
\int_{0}^{u_{1}} d\mbox{u} A \left( u \right) 
\left[
1 + \dfrac{1}{n}
\left(
\ln P_{0} +
\int _{0}^{u} d\mbox{u'} A \left( u' \right)
\right)
\right]
\left[
\ln P_{0} +
\int _{0}^{u} d\mbox{u'} A \left( u' \right)
\right]
^n
\end{equation}

\begin{equation}
\label{computingP0}
P_{0}  = \frac{1}{D \int _{0} ^{u_{1}} d \mbox{u} \exp \left[ \int _{0} ^{u} d \mbox{u'} A \left( u' \right) \right]}
\end{equation}

\begin{equation}
\label{computingu1}
\int_{0}^{u_{1}} d \mbox{u} A \left( u \right) = 1
\end{equation}

where $ \Pi_{z = 0} \equiv \Pi_{q' = 1} = \Pi_{q = 1} $, $ z \equiv q' - 1 = 1 - q$ and $ 0 \leq z < 1 $ as $ 0 < q \leq 1 $. Together, equations \eqref{Taylor}, \eqref{coeffs}, \eqref{computingP0} and \eqref{computingu1} allow computation of $ \Pi_{z} $ (hence, of $ \Pi_{q'} $) once $ A \left( u \right) $ and $ D $ are known.

\subsection{Stable distributions of probability.} The evolution of the system is a succession of relaxed states, whose nature depends on $ J $. If $ J = 0 $ then Gibbs' statistics holds and the probability distribution of the relaxed state is a Boltzmann exponential. 

For weakly dissipating systems $ \vert J \vert \neq 0 $ is small and each relaxed state is approximately an equilibrium ($ S_{z} = \max $, hence $ \widehat{S_{z}} = \max $). It follows that \eqref{Pq} describes the probability distribution and that 
small $ \kappa $ fluctuations near a relaxed state are Gaussian distributed with variance $ \left( \frac{\partial^2 \widehat{S_{z}} }{\partial \kappa^2}\right)^{-1} $.

Now, stability requires that fluctuations, once triggered, relax back to the initial equilibrium state, sooner or later. In other words, the larger the variance the larger the fluctuations of $ \kappa $ which the relaxed state is stable against. 
Accordingly, the relaxed state which is stable against the fluctuations of $ \kappa $ of largest variance corresponds to $ \frac{\partial^2 \widehat{S_{z}} }{\partial \kappa^2} = 0 $. Then, \eqref{monotonicdependence} gives $ \frac{\partial^2 S_{z} }{\partial \kappa^2} \propto \frac{\partial^2 \widehat{S_{z}} }{\partial \kappa^2} = 0 $.

Arbitrariness in the definition of $ \kappa $ allows us to identify it with a perturbation of $ z $ (so that $ d \kappa = d z $). 
Moreover, for small $ \vert J \vert $ (i.e., negligible entropy flux across the boundary) equation \eqref{balanceofSq} makes any increase of $ S_{z} $ in a given time interval $ \Delta t $ to be equal to $ \Delta t \cdot \Pi_{z} $, so that $ \Delta t \cdot \frac{d^2 \Pi_{z}}{dz^2} = \frac{\partial^2 S_{z}}{\partial z^2} $. 
Thus, the 'most stable probability distribution' (i.e., the probability distribution which is stable against the fluctuations of $ z $ of largest amplitude) is given by \eqref{Pq} with $ q = 1 - z_{c}$ and $ z_{c} $ such that:

\begin{equation}
\label{stability}
\left( \dfrac{d^2 \Pi_{z} }{d z^2} \right)_{z = z_{c}} = 0 
\quad \mbox{;} \quad
0 < z_{c} < 1
\end{equation}

According to \eqref{stability}, $ \frac{d \Pi_{z} }{d z} $ achieves an extremum value at $ z = z_{c} $. In order to ascertain if this extremum is a maximum or a minimum, we recall that the change $ dS_{z} = dz \cdot \left( \frac{dS_{z}}{dz} \right)_{z_{c}} = dz \cdot \Delta t \cdot \left( \frac{d \Pi}{dz} \right)_{z_{c}} $ in $ S_{z} $ due to a fluctuation $ dz $ of $ z $ around $ z = z_{c} $ occurring in a time interval $ \Delta t $ is $ \geq 0 $ (this is true for $ S_{z = 0} $ as fluctuations involve irreversible processes and for $ S_{z \neq 0} $ as \eqref{GEC} describes relaxation the same way regardless of the value of $ z $); it achieves its minimum value $ 0 $ if $ J = 0 $ and the relaxed state is a true equilibrium ($ S_{z = z_{c}} = \max $, $ \left( \frac{dS_{z}}{dz} \right)_{z_{c}} \propto \left( \frac{d \Pi}{dz} \right)_{z_{c}} = 0 $). In the weak dissipation limit the structure of the relaxed state is perturbed only slightly, and we may still reasonably assume $ \left( \frac{d \Pi}{dz} \right)_{z_{c}} = \min $ even if its value $ \propto O \left( \vert J \vert^2 \right) $ is $ \neq 0 $ (again, \eqref{GEC} acts the same way). In agreement with \eqref{stability}, we obtain:

\begin{equation}
\label{stability2}
\left( \dfrac{d^3 \Pi_{z} }{d z^3} \right)_{z = z_{c}} > 0 
\end{equation}

Remarkably, \eqref{stability} and \eqref{stability2} hold regardless of the actual relaxation time of the fluctuation; it applies therefore also to the slow dynamics of $ z \left( t \right) $. We stress the point that \eqref{stability} and \eqref{stability2} have no means been said to ensure the actual existence of a relaxed configuration in the weak dissipation limit with something like a most stable probability distribution. But if such a thing exists, then it behaves as a power law with exponent $ \frac{1}{1 - q} = z_{c}^{-1}$ if $ 0 < z_{c} < 1 $, where $ z_{c} $ corresponds to a minimum of $ \frac{d \Pi_{z}}{dz} $ according to \eqref{stability} and \eqref{stability2}. We rule out Boltzmann exponential in this case; indeed, if the power law is stable at all then it is stable against larger fluctuations than the Boltzmann exponential, because fluctuations are larger for $ 0 < z <1 $ than for $ z = 0 $ \cite{VivesandPlanes}. Together, \eqref{Taylor}, \eqref{coeffs}, \eqref{computingP0}, \eqref{computingu1}, \eqref{stability} and \eqref{stability2} provide us with $ z_{c} $ once $ A \left( u \right) $ and $ D $ are known. We discuss an application below.

\section{The impact of noise on 1D maps.}
\label{SEC5bis}

\subsection{Boltzmann vs. power law.} We apply the results of Sec. \ref{SEC5} to the description of the impact of noise on maps. Let us introduce a discrete, autonomous, one-dimensional map 

\begin{equation}
\label{map}
Q_{i + 1} =  G \left( Q_{i} \right)
\end{equation}

where $ i = 0, 1, 2, \ldots $, $ Q_{i} \geq 0 $ for all $ i $'s, the initial condition $ Q_{i = 0} =  Q_{0} $ is known and $ G $ is a known function of its argument. 
If the system evolves all along a time interval $ \tau $, then \eqref{map} leads to the differential equation $ \frac{dx \left( t' \right)}{dt'} = A \left( x \right) $ 
, provided that we 
define $ A \left( x \right) \equiv  G \left( x \right) - x $, 
identify $ Q_{i + 1} $ and $ Q_{i} $ with $ \frac{x \left( t' + \Delta t' \right) }{\Delta t'} $ and $ \frac{x \left( t' \right) }{\Delta t'} $ respectively, write $ t' = i \cdot \Delta t' $ and consider a time increment $ \Delta t' \ll \tau $ ('continuous limit'). We suppose $ A \left( x \right) $ to be integrable and well-behaved at infinity (see Sec. \ref{SEC5}).

Map \eqref{map} includes no noise. The latter may be either additive or multiplicative, and may affect $ Q_{i} $ at any 'time' $ i $; for instance, it may perturb $ Q_{0} $. In the continuous limit, we introduce a distribution function  $ P \left( x , t \right) $ such that $ P \left( x , t \right) dx $ is the probability of finding the coordinate in the interval between $ x $ and $ x + dx $ at the time $ t \equiv \eta \cdot t' $. We discuss the role of the constant $ \eta > 0$ below. In order to describe the impact of noise, we modify the differential equation above as follows: 

\begin{equation}
\label{stochastic}
\eta \dfrac{dx}{dt} = A \left( x \right) + h \left( x , t \right) \zeta \left( t \right)
\quad \mbox{where} \quad
A \left( x \right) \equiv  G \left( x \right) - x
\end{equation}

where $ h \left( x , t \right) \propto P \left( x , t \right) ^{\frac{z}{2}} $, $ 0 \leq z < 1 $, the noise $ \zeta \left( t \right) $ satisfies $ \left\langle \zeta \right\rangle = 0 $ and $ \left\langle \zeta \left( t \right) \zeta \left( t' \right) \right\rangle = 2 \eta D \delta \left( t - t' \right) $, and the brackets $ \left\langle \right\rangle $ denote time average \cite{Casas}. We leave the value of $ z $ unspecified.

According to the discussion of equation (18) of \cite{Borland}, \eqref{stochastic} is associated with \eqref{FP} which rules $ P \left( x , t \right) dx $. The choice of $ D $ in \eqref{FP} is equivalent to the choice of the noise level in \eqref{stochastic}, and our assumption that $ D $ is constant and uniform in Sec. \ref{SEC5} means just that the noise level is the same throughout the system at all times. 
Well-behavedness of $ A $ at infinity allows H-theorem to apply to \eqref{FP}. Let a relaxed state exist as an outcome of the evolution described by \eqref{FP}. We know nothing about $ \vert J \vert $. For the moment, let us discuss the case $ J \neq 0 $. 

If $ J \neq 0 $ then the definition of $ J $ in \eqref{FP} allows us to choose a value $ \eta $ large enough that the weak dissipation limit of small $ \vert J \vert $ applies. 
Once the dynamics (i.e., the dependence of $ G $ on its own argument) and the noise level (i.e., $ D $) are known, then \eqref{Taylor}, \eqref{coeffs}, \eqref{computingP0} and \eqref{computingu1} allow us to compute the value $ z_{c} $ of $ z $ which minimizes $ \frac{d \Pi_{z}}{dz} $. According to our discussion of  \eqref{stability} and \eqref{stability2} in Sec. \ref{SEC5}, if such minimum exists and $ 0 < z_{c} < 1 $ then the most stable probability distribution is power-like with exponent $ \frac{1}{z_{c}} $. Otherwise, no relaxed state may exist for $ J \neq 0 $. If, nevertheless, such state exists, then $ J = 0 $ and its probability distribution is a Boltzmann exponential. 

We have just written down a criterion to ascertain whether the outcome $ Q_{i} $ of a map \eqref{map} follows an exponential or a power law distribution function as $ i \rightarrow \infty $ (and provided that such a distribution function may actually be defined in this limit) whenever the map is affected by noise, and regardless of the nature (additive or multipicative) of the latter. Only the map dynamics $ G \left( Q_{i} \right) $ and the noise level $ D $ are required. In contrast with conventional treatment, no numerical solution of \eqref{map} is required, and, above all, $ q $ is no \textit{ad hoc} input anymore. This is the criterion looked for in Sec. \ref{SEC1}.

\subsection{An example.} As an example, we consider the map \eqref{map} where 

\begin{equation}
\label{Sanchez}
G \left( x \right) = r x \exp \left( - \vert 1 - a \vert x \right)
\end{equation}

and $ r $ and $ a $ are real, positive numbers. For typical values $ a = 0.8 $ and $ 1 < r < 7 $ equations \eqref{map} and \eqref{Sanchez} are relevant in econophysics, where $ x $ and $ P $ are the wealth and the distribution of richness respectively. We refer to Ref.  \cite{Sanchez} - and in particular to its equation (2) - where noise is built in to the initial conditions (which are completely random), and after a transient the system relaxes to a final, asymptotical state, left basically unaffected by fluctuations. 

\begin{figure}[!h]
\centering
\includegraphics[scale=0.5]{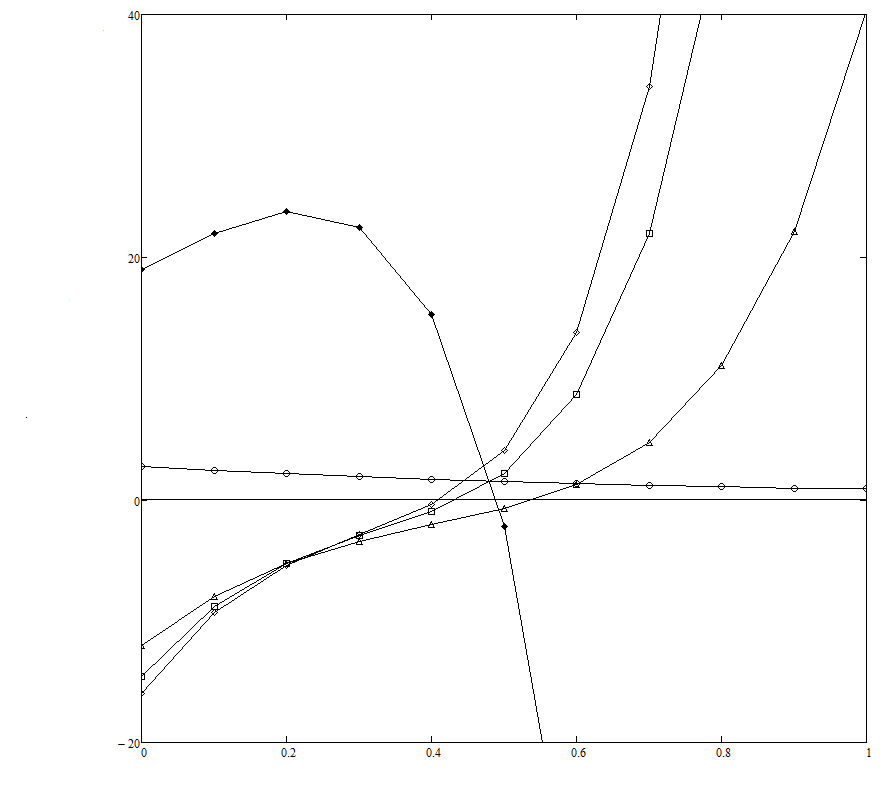}
\caption{\textit{$ \frac{d^2 \Pi_{z} }{d z^2} $ (vertical axis) vs. $ z $ (horizontal axis) for $ r = 1/3 $ (black diamonds), $ r = 1/2 $ (empty circles), $ r = 2 $ (triangles), $ r = 4 $ (squares), $ r = 6 $ (empty diamonds). In all cases $ D = 0.1 $.  If $ r = 1/3 $ then the slope of the curve at the point $ z = z_{c} $ where it crosses the $ \frac{d^2 \Pi_{z} }{d z^2} = 0 $ axis is negative, i.e. \eqref{stability2} is violated. If $ r = 1/2 $ then $ z_{c} $ lies outside the interval $ 0 \leq z_{c} < 1 $, i.e. \eqref{stability} is violated. Both \eqref{stability} and \eqref{stability2} are satisfied for $ r = 2 $ (with $ z_{c} = 0.452 $), $ r = 4 $ (with $ z_{c} = 0.438 $) and $ r = 6 $ (with $ z_{c} = 0.412 $).}}
\label{Fig1.png}
\end{figure}

\begin{figure}[!h]
\centering
\includegraphics[scale=0.5]{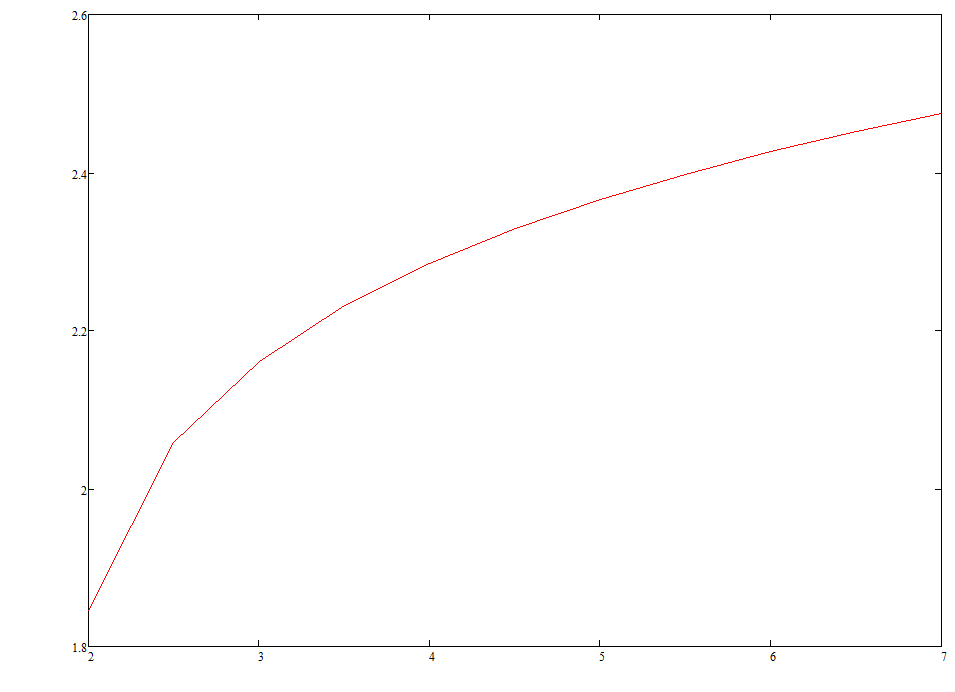}
\caption{\textit{Exponent $ \frac{1}{z_{c}} $ (vertical axis) of the power-law distribution function of the relaxed state vs. $ r $ (horizontal axis).}}
\label{Fig2.png}
\end{figure}

\begin{figure}[!h]
\centering
\includegraphics[scale=0.5]{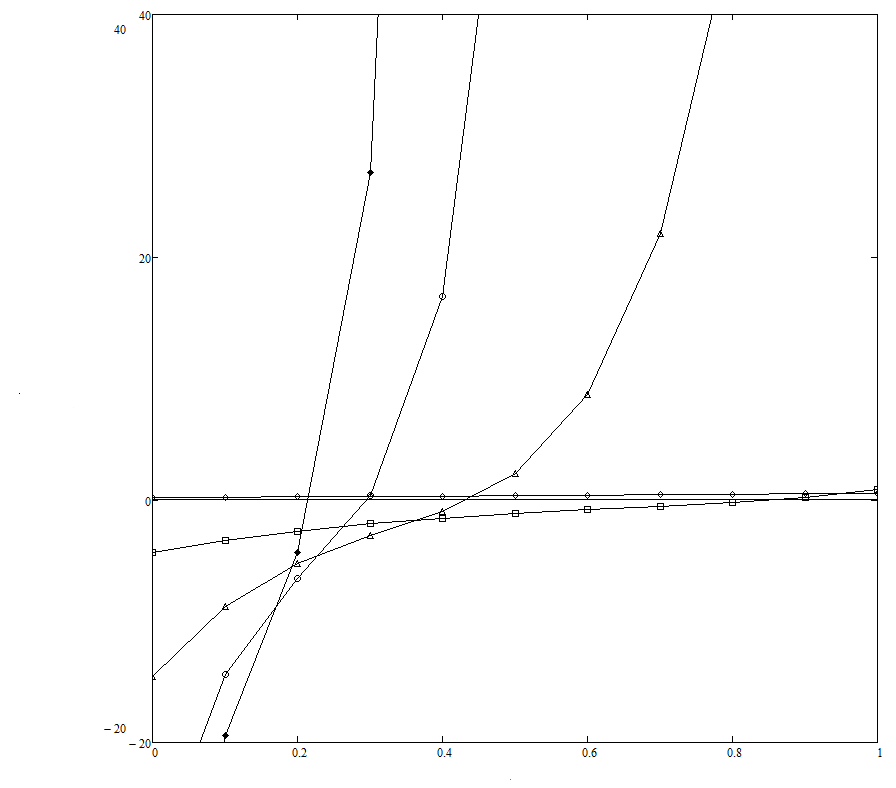}
\caption{\textit{$ \frac{d^2 \Pi_{z} }{d z^2} $ (vertical axis) vs. $ z $ (horizontal axis) for $ D = 0.001 $ (black diamonds), $ D = 0.01 $ (empty circles), $ D = 0.1 $ (triangles), $ D = 2 $ (squares), $ D = 10 $ (empty diamonds). In all cases $ r = 4 $. Even if a relaxed state exists, the  larger $ D $, the stronger the noise, the nearer $ z_{c} $ to the bounds of the interval $ \left[ 0 , 1 \right) $. If $ D > 1 $ then $ z_{c} $ does not belong to the interval, and Boltzmann's exponential distribution rules the relaxed state.}}
\label{Fig3.png}
\end{figure}

We assume $ a = 0.8 $ everywhere in the following. Fig. \ref{Fig1.png} displays $ \frac{d^2 \Pi_{z} }{d z^2} $ (normalized to $ 2J $) vs. $ z $ as computed from \eqref{Taylor}, \eqref{coeffs}, \eqref{computingP0}, \eqref{computingu1} and \eqref{Sanchez} at various values of $ r $ and with the same value of $ D \left( = 0.1 \right) $. When dealing with \eqref{Taylor} we have taken into account powers of $ z $ up to $ z^7 $ (included). We performed all algebraic computations and definite integrals with the help of MATHCAD software. If $ r \leq 1 $ then no $ z_{c} $ is found which satisfies both \eqref{stability} and \eqref{stability2}, so that Boltzmann distribution is expected to describe the asymptotic dynamical state of the system. This is far from surprising, as only damping counteracts noise in \eqref{stochastic} in the $ r \rightarrow 0$ limit, like in a Brownian motion. In contrast, if $ r > 1 $ then all $ z_{c} $'s lies well inside the interval $ 0 < z_{c} < 1$, and a power law is expected to hold; the correponding exponential depends on $ r $ only weakly - see Fig. \ref{Fig2.png} - as the values of the $ z_{c} $'s are quite near to each other. Finally, Fig. \ref{Fig3.png} displays how $ z_{c} $ depends on $ D $ (i.e., the noise level) at fixed $ r $; it turns out that noise tends to help relaxation to Boltzmann's distribution, as expected.

Our results seem to agree with the results of the numerical simulations reported in \cite{Sanchez}. If $ r < 1 $ then the average value of $ x $ relaxes to zero (just as predicted by standard analysis of \eqref{map} in the zero-noise case) and random fluctuations occur. In contrast, if $ r > 1 $ then the typical amplitude of the fluctuations is much larger; nevertheless, a distribution function is clearly observed, which exhibits a distinct power-law, Pareto-like behaviour. The exponent is 2.21, in good agreement with the values displayed in our Fig. \ref{Fig2.png}. (The exponent in Pareto's law is $ \approx $ 2.15). We stress the point that we have obtained our results with no numerical solution of \eqref{map} and with no postulate concerning non-extensive thermodynamics, i.e. no assumption on $ q $.

\section{Conclusions}
\label{SEC6} 

Gibbs' statistical mechanics describes the distribution of probabilities of the microstates of (grand-)canonical systems at thermodynamical equilibrium with the help of Boltzmann's exponential. In contrast, this distribution follows a power law in stable, steady ('relaxed') states of many physical systems. With respect to a power-law-like distribution, non-extensive statistical mechanics \cite{TsallisORIG} \cite{TSALLISMENDESPLASTINO} formally plays the same role played by Gibbs' statistical mechanics with respect to Boltzmann distribution: a relaxed state corresponds to a constrained maximum of Gibbs' entropy and to its generalization $ S_{q} $ in Gibbs' and non-extensive statistical mechanics respectively. Generalization of some  results of Gibbs' statistical mechanics to non-extensive statistical mechanics is available; the latter depends on the dimensionless quantity $ q $ and reduces to Gibbs' statistical mechanics in the limit $ q \rightarrow 1 $, just like $ S_{q} $ reduces to Gibbs' entropy $ S_{q = 1} $ in the same limit. The quantity $ q $ measures the lack of additivity of $ S_{q} $ and provides us with the slope of the power-law-like distribution, the Boltzmann distribution corresponding to $ q = 1 $: $ S_{q} $ is an additive quantity if and only if $ q = 1 $. 

The overwhelming success of Gibbs' statistical mechanics lies in its ability to provide predictions (like e.g. the positivity of the specific heat at constant volume) even when few or no information on the detailed dynamics of the system is available. Stability provides us with an example of such predictions. According to Einstein's formula,  deviations from thermodynamic equilibrium which lead to a significant reduction of Gibbs' entropy ($ \Delta S_{q = 1} < 0$) have vanishing small probability $ \propto \exp \left( \Delta S_{q = 1} \right) $. In other words, significant deviations of the probability distribution from the Boltzmann exponential are exponentially unlikely in Gibbs' statistical mechanics.

Moreover, additivity of $ S_{q = 1} $ and of other quantities like the internal energy allows to write all of them as the sum of the contributions of all the small parts the system is made of. If, furthermore, every small part of a physical system corresponds locally to a maximum of $S_{q = 1}$ ('local thermodynamical equilibrium, LTE) at all times during the evolution of the system, then this evolution is bound to satisfy the so-called 'general evolution criterion' (GEC) \cite{GlansdorffPrigogine}, an inequality involving total time derivatives of thermodynamical quantities which follows from Gibbs-Duhem equation. In particular, GEC applies to the relaxation of perturbations of a relaxed state of the system, if any such state exists.

In contrast, lack of \textit{a priori} knowledge of $ q $ limits the usefulness of non-extensive statistical mechanics; for each problem, such knowledge requires either solving the detailed equations of the dynamics (e.g., the relevant kinetic equation ruling the distribution probability of the system of interest) or performing \textit{a posteriori} analysis of experimental data, thus reducing the attractiveness of non-extensive statistical mechanics.

However, it is possible to map non-extensive statistical mechanics into Gibbs' statistical mechanics \cite{TsallisORIG} \cite{VivesandPlanes} \cite{Abe}. A quantity $\widehat{S_{q}}$ exists which is both additive and monotonically increasing function of $ S_{q} $ for arbitrary $ q $. Thus, relaxed states of non-extensive thermodynamics correspond to $\widehat{S_{q}} = \max$, and additivity of $\widehat{S_{q}}$ allow suitable generalization of both Einstein's formula and Gibbs-Duhem equation to $ q \neq 1 $ \cite{VivesandPlanes}, which in turn ensure that strong deviations from this maximum are exponentially unlikely and that LTE and GEC still hold, formally unaffected, in the $ q \neq 1 $ case respectively. 

These generalizations allow thermodynamics to provide an unified framework for the description of both the relaxed states (via Einstein's formula) and the relaxation processes leading to them (via GEC) regardless of the value of $ q $, i.e. of the nature  - power law ($ q \neq 1 $) vs. Boltzmann exponential ($ q = 1 $) - of the probability distribution of the microstates in the relaxed state. 

For further discussion we have focussed our attention on the case of a continuous, one-dimensional system described by a nonlinear Fokker Planck equation \cite{Casas}, where the impact of a driving force is counteracted by diffusion (with diffusion coefficient $ D $). It turns out that it is the the interaction with the external world which allows the probability distribution in the relaxed states to differ from a Boltzmann's exponential. Moreover, Einstein's formula in its generalized version implies that the value $ z_{c} $ of $ z \equiv 1 - q $ of the 'most stable probability distribution' (i.e., the probability distribution of the relaxed state which is stable against fluctuations of largest amplitude) corresponds to a minimum of $ \frac{d\Pi_{z}}{dz} $, $ \Pi_{z} \cdot dt $ being the amount of $ S_{z} $ produced in a time interval $ dt $ by irreversible processes occurring in the bulk of the system. Finally, if a relaxed state exists and $ 0 < z_{c} < 1 $, then the most stable probability distribution is a power law with exponent $ \frac{1}{z_{c}} $; otherwise, it is a Boltzmann exponential. Since $ \Pi_{z} $ depends just on $ z $, $ D $ and the driving force, the value of $ z_{c} $ - i.e., the selection of the probability distribution - depends on the physics of system only (i.e. on the diffusion coefficient and the driving force): \textit{a priori} knowledge of $ q $ is required no more.
 
We apply our result to the Fokker Planck equation associated to the stochastic differential equation obtained in the continuous limit from a one-dimensional, autonomous,  discrete map affected by noise. Since no assumption is made on $ q $, the noise may be either additive or multiplicative, and the Fokker Planck equation may be either linear or nonlinear. If the system evolves towards a system which is stable against fluctuations then we may ascertain if a power-law statistics describes such state - and with which exponent - once the dynamics of the map and the noise level are known, without actually computing many forward orbits of the map. 

As an example, we have analyzed the problem discussed in \cite{Sanchez}, where a particular one-dimensional discrete map affected by noise leads to an asymptotic state described by a Pareto-like law for selected values of a control parameter. Our results agree with those of \cite{Sanchez} as far as both the exponent of the power law and the range of the control parameter, with the help of numerical simulation of the dynamics and of no assumption about $ q $.

Extension to multidimensional maps will be the task of future work.

\vspace{6pt} 
%

\section*{Acknolwledgments}

Useful discussions and warm encouragement with Prof. W. Pecorella, Università di Tor Vergata, Roma, Italy are gratefully acknowledged.

\appendix
%
\section{Non-existence of $ s $ for $ q \neq 1 $}
\label{AA}

If $ q \rightarrow 1 $ then $ S_{q'} \rightarrow S_{q = 1} = - \int d\mbox{x} P \ln P $, which leads immediately to $ S_{q = 1} = \int d\mbox{x} \rho s $ with $ \rho \propto P $ and $ s \propto \ln P $. Let us suppose that a similar expression holds for $ S_{q} $ and $ q \neq 1 $. Of course, $ \rho $ does not depend on $ q $, so we keep $ \rho \propto P $ even if $ q \neq 1 $. In the latter case, \eqref{Sq} gives $ S_{q'} \propto \int d\mbox{x} \rho \frac{1 - P^{q' - 2}}{q' - 1} $, and agreement of \eqref{def0} with \eqref{Sq} requires that we identify $ \sum_{k} \left( p_{k} \right) ^{q'} $ and $ \int  d\mbox{x} P^{q' - 2} $. However, this identification is possible at no value of $ q' $, because the normalization conditions $ \sum_{k} p_{k} = 1 $ and $ \int d\mbox{x} P = 1 $ make the $ p_{k} $'s and $ P $ to transform like $ p_{k} \rightarrow p_{k} \cdot k^{-1}$ and $ P \rightarrow P \cdot k^{-1} $ respectively under the scaling transformation $ x \rightarrow k x $, so that $ \sum_{k} \left( p_{k} \right) ^{q'} \rightarrow k^{-q'} \cdot \sum_{k} \left( p_{k} \right)^{q'} $ and $ \int  d\mbox{x} P^{q' - 2} \rightarrow k^{2 - q'} \cdot \int  d\mbox{x} P^{q' - 2}$. Identification of $ \sum_{k} \left( p_{k} \right) ^{q'} $ and $ \int  d\mbox{x} P^{q' - 2} $ is impossible because these two quantities behave differently under the same scaling transformation; then, no definition of $ s $ is self-consistent unless $ q = 1 $. 

This result follows from the non-additivity of $ S_{q} $: if $ q \neq 1 $ then the entropy of the whole system is not the sum of the entropies of the small masses the systems is made of. Physically, this suggests that strong correlations exixt among such masses; indeed, strongly correlated variables are precisely the topic which non-extensive statistical mechanics is focussed on \cite{Umarov}. 

Accordingly, straightforward generalization of LTE and GEC with the help of $ S_{q} $ to the $ q \neq 1 $ case is impossible. This is why we need the additive quantity $ \widehat{S_{q}} $ in order to build a local thermodynamics, and to generalize the results of Sec. \ref{SEC3} in Sec. \ref{SEC4}. As discussed in the text, the results of Sec. \ref{SEC4} may involve $ \Pi_{q} $ rather than $ \widehat{\Pi_{q}} $ just because $ \frac{d\widehat{S_{q}}}{dS_{q}} > 0$.

\section{Proof of \eqref{Taylor}, \eqref{coeffs}, \eqref{computingP0} and \eqref{computingu1}}
\label{BB}

We derive from both \eqref{uniformJ}, \eqref{Piq} and the definition of $ J $ the Taylor-series development \eqref{Taylor} of $ \Pi_{z} = \Pi_{q'} $ in powers of $ z $, centered in $ z = 0 $:

%
\begin{equation}
\label{Taylor0}
\Pi_{z} = \Pi_{z = 0} + \sum _{n = 1} ^{\infty} a_{n} z^n
\quad \mbox{;} \quad
a_{n} = \dfrac{\left( -1 \right)^{n - 1}}{\left( n - 1 \right)!} \left( 2 J \right) \int d\mbox{u} \dfrac{d \ln P_{J, q = 1}}{d u} \left( 1 + \dfrac{\ln P_{J, q = 1}}{n} \right) \left( \ln P_{J, q = 1} \right)^{n - 1}
\end{equation}

where $ u \equiv \frac{x}{D} $ and $ P_{J, q = 1} = P_{J, q = 1} \left( u \right)$ is the $ q' = q = 1 $ solution for arbitrary $ J $ with the boundary condition $ P_{J, q = 1} \left( u = u_{0} \right) = P_{0} $ of \eqref{FP}. 
Starting from \eqref{FP} and \eqref{Pq} (for $ q = 1 $), the method of variation of constants gives: 

\begin{equation}
\label{Pjq1}
P_{J, q = 1} \left( u \right) = 
\left\lbrace
P_{0} 
- \eta J \int _{u_{0}} ^{u} d \mbox{u'} 
\exp \left[ 
- \int _{u_{0}} ^{u'} d \mbox{u''} A \left( u'' \right) 
\right] 
\right\rbrace 
\exp \left[ 
\int  _{u_{0}} ^{u} d \mbox{u'} A \left( u' \right) 
\right]
\end{equation}

The normalization condition holds

\begin{equation}
\label{normalizationcondition}
1 = 
\int d \mbox{x} P_{J, q = 1} \left( x \right)
= D \int d \mbox{u} P_{J, q = 1} \left( u \right)
\end{equation}

where $ P_{J, q = 1} $ is approximately given by \eqref{Pq} in weakly dissipating systems. We assume $ x \geq 0 $ and $ u_{0} = 0 $ with no loss of generality (originally, $ x_{1k} $ is an energy). We define $ u_{1} $ such that $ P_{0}
= \eta J \int _{u0} ^{u_{1}} d \mbox{u'} 
\exp \left[ 
- \int _{0} ^{u'} d \mbox{u''} A \left( u'' \right) 
\right] $. It is unlikely that $ \beta U \gg 1 $; thus, we approximate \eqref{Pjq1} as :

\begin{equation}
\label{approximation}
P_{J, q = 1} = P_{J, q = 1} \left( 0 \right) \exp \left( \int _{0} ^{u} d \mbox{u'} A \left( u' \right) \right) \quad \mbox{for} \quad 0 \leq u \leq u_{1} \quad \mbox{;} \quad P_{J, q = 1} =  0 \quad \mbox{otherwise}
\end{equation}

According to \eqref{approximation}, the domain of integration in \eqref{Taylor}, \eqref{Pjq1} and \eqref{normalizationcondition} reduces to $ 0 \leq u \leq u_{1} $. Thus, \eqref{Taylor} and \eqref{approximation} lead to \eqref{coeffs}, and  \eqref{normalizationcondition} and \eqref{approximation} lead to \eqref{computingP0}.  Finally, \eqref{uniformJ}, \eqref{balanceofSq} and \eqref{approximation} lead to: $ \Pi_{q'} = J \int_{0} ^{u_{1}} d\mbox{u} A $ in relaxed state, while \eqref{uniformJ}, \eqref{Piq} and \eqref{approximation} give: $ \Pi_{q'} = \frac{\eta \vert J \vert^2}{P_{J, q = 1} \left( 0 \right)} \int_{0}^{u_{1}} d \mbox{u} \exp \left( - \int_{0}^{u} d \mbox{u'} A \right) $. After eliminating $ \Pi_{q'} $ and replacing the definition of $ u_{1} $ we obtain \eqref{computingu1}.
%



\begin{thebibliography}{widest-label}

\bibitem{TsallisORIG}
  C. Tsallis, J. Stat. Phys. \textbf{52} 1/2, 479 (1988)

\bibitem{TSALLISMENDESPLASTINO}
  C. Tsallis, R. S. Mendes, A. Plastino, Physica A \textbf{261} 534 (1998)  

\bibitem{VivesandPlanes}
  E. Vives, A. Planes, PRL \textbf{88} 2, 020601 (2002)
  
\bibitem{Landau}
  L. D. Landau, E. M. Lifshitz, \textit{Statistical Physics} Pergamon Oxford (1959)

\bibitem{Tsallis2}  
  C. Tsallis, C. Anteneodo, L. Borland, R. Osorio, Physica A: Statistical Mechanics and its Applications \textbf{324}, 1 pages 89-100 (2003)
  
\bibitem{Abe}
  S. Abe, Physica A \textbf{300} 417–423 (2001)
  
\bibitem{Marino}
  M. Marino, Physica A: Statistical Mechanics and its Applications \textbf{386}.1 (2007): 135-154   
  
\bibitem{PrigogineDefay}  
  I. Prigogine, R. Defay, \textit{Chemical Thermodynamics} Longmans-Green, London, (1954)

\bibitem{GlansdorffPrigogine} 
  P. Glansdorff, I. Prigogine, Physica \textbf{30} 351 (1964)

\bibitem{Di Vita}    
  A. Di Vita, Phys. Rev. E \textbf{81} 041137 (2010)
%
  
\bibitem{Casas}    
  G. A. Casas, F. D. Nobre, E. M. F. Curado, Phys. Rev. E \textbf{86} 061136 (2012)

\bibitem{Tatsuaki}      
  W. Tatsuaki, Physics Letters A Volume \textbf{297}, Issues 5–6, 20 May 2002, Pages 334–337
  
\bibitem{Khuntia}      
  A. Khuntia, P. Sahoo, P. Garg, R. Sahoo, J. Cleymans, \textit{Speed of Sound in a System Approaching Thermodynamic Equilibrium} Proceedings of the DAE-BRNS Symp. on Nucl. Phys. \textbf{61} (2016)
  
\bibitem{Borland}  
  L.Borland, Phys. Rev E \textbf{57}, 6, 6634

\bibitem{Haubold}    
  H. J. Haubold, A. M., Mathai, R. K. Saxena, Astrophysics and Space Science, \textbf{290}(3), 241-245 (2004)
 
\bibitem{Ribeiro}
  M. S. Ribeiro, F. D. Nobre, E. M. F. Curado, Phys. Rev. E \textbf{85}, 021146 (2012)  
    
\bibitem{Wedemann}
  R. S. Wedemann, A. R. Plastino, C. Tsallis, Phys. Rev. E \textbf{94}, 062105 (2016) 
  
\bibitem{Umarov}
  S. Umarov, C. Tsallis, S. Steinberg, Milan J. of Mathematics \textbf{76}.1 (2008): 307-328  
  
\bibitem{Wada}   
  T. Wada, A. M. Scarfone, Physics Letters A \textbf{335}.5, 351-362 (2005)

\bibitem{Naudts}   
  J. Naudts, Physica A: Statistical Mechanics and its Applications \textbf{340}.1, 32-40 (2004)
  
\bibitem{Sanchez}     
   J. R. Sánchez, R. Lopez-Ruiz, The European Physical Journal-Special Topics \textbf{143}.1 (2007): 241-243.

\end{thebibliography}
\end{document}